\documentclass{webofc}
\usepackage[varg]{txfonts}   
\usepackage{natbib}
\usepackage{hyperref}
\usepackage{xcolor}
\usepackage{sidecap}
\newcommand{\cl}{CutLang }
\newcommand{\atnm}{adl2tnm }

\usepackage[justification=centering]{caption}

\begin{document}
\title{Recent advances in \texttt{ADL}, \texttt{CutLang} and \texttt{adl2tnm}}
%
%

\author{\firstname{Harrison B.} \lastname{Prosper}\inst{1}\fnsep\thanks{\email{harry@hep.fsu.edu}} \and
\firstname{Sezen} \lastname{Sekmen}\inst{2}\fnsep\thanks{\email{ssekmen@cern.ch}} \and
\firstname{Gokhan} \lastname{Unel}\inst{3}\fnsep\thanks{\email{gokhan.unel@cern.ch}} \and
\firstname{Arpon} \lastname{Paul}\inst{4}\fnsep\thanks{\email{apaul@ictp.it}} }
\institute{Department of Physics, Florida State University, Tallahassee, FL, USA
\and
Center for High Energy Physics, Kyungpook National University, Daegu, South Korea
\and
Physics and Astronomy Department, University of California at Irvine, Irvine, CA, USA
\and
The Abdus Salam International Centre for Theoretical Physics, Trieste, Italy
}
\abstract{%
This paper presents an overview and features of an Analysis Description Language (ADL) designed for HEP data analysis.  ADL is a domain-specific, declarative language that describes the physics content of an analysis in a standard and unambiguous way, independent of any computing frameworks.  It also describes infrastructures that render ADL executable, namely CutLang, a direct runtime interpreter (originally also a language), and adl2tnm, a transpiler converting ADL into C++ code. In ADL, analyses are described in human-readable plain text files, clearly separating object, variable and event selection definitions in blocks, with a syntax that includes mathematical and logical operations, comparison and optimisation operators, reducers, four-vector algebra and commonly used functions. Recent studies demonstrate that adapting the ADL approach has numerous benefits for the experimental and phenomenological HEP communities.  These include  facilitating the abstraction, design, optimization, visualization, validation, combination, reproduction, interpretation and overall communication of the analysis contents and long term preservation of the analyses beyond the lifetimes of experiments.  Here we also discuss some of the current ADL applications in physics studies and future prospects based on static analysis and differentiable programming.

}
\maketitle
\section{Introduction}

High energy physics (HEP) experiments are collecting unprecedented amounts of data.  In order to explore these data for hints of new physics, or to perform high precision measurements, physicists are designing an ever growing number of elaborate analyses.  The physics content of these analyses consists of defining objects and variables used for classifying events as signal or background, selecting events, re-weighting simulated events to improve their agreement with real events, estimating backgrounds, and interpreting experimental results by comparing them to theory predictions. These tasks are traditionally performed using analysis software frameworks that organize the tasks into a computational pipeline. The frameworks integrate a diverse set of operations from data access to event selection, from histogramming to statistical analysis.  These frameworks are written in general purpose languages (GPLs) like C++ or Python. At the Large Hadron Collider (LHC) at CERN, many analysis teams have their own dedicated frameworks. There are also frameworks such as   CheckMate~\cite{CheckMATE,KIM2015535,2016chep.confE.120T} and MadAnalysis~\cite{MadAnalysis,Conte_2014} for phenomenology studies and the Rivet framework~\cite{Waugh:2006ip,BUCKLEY20132803} mainly focused on unfolded LHC measurements.

Mastering such frameworks requires a high level knowledge of both GPLs and software architectures. These requirements erect a barrier between data and the physicist who may simply wish to try an analysis idea. An extra challenge is having the physics content scattered throughout different components of the framework code  makes implementing and working with different physics ideas less straightforward and efficient, even for experienced analysts.  Working with multiple frameworks is a further issue, since each framework has a different way of implementing the physics content.  Yet, attempting to build a single, unified framework for everyone is an unrealistic goal.

An alternative approach to address these issues is to use a domain specific language (DSL) for describing the physics contents.  In this report, we present the developments on such a language, called an {\bf Analysis Description Language (ADL)} that can fully and unambiguously describe the complete physics algorithm of a HEP analysis in a framework-independent manner.  Here, framework-independence means that ADL is not part of a single framework, but can be used by any tool or framework that can parse and execute its syntax.  ADL is also declarative, meaning that it expresses the analysis logic without describing its control flow.  This approach has many advantages: ADL can make analysis writing significantly easier by eliminating coding complexities, thus enabling analysts with different computing skill levels to focus on analysis design.  A standard language tailored to HEP will also help self-documentation and easier communication of the physics content within the analysis team, with referees, between different experiments, between experimentalists and theorists, etc., which will make analysis validation and review easier.  Moreover, ADL will directly facilitate the (re)interpretation of analysis results for both experimentalists and theorists~\cite{Abdallah:2020pec}.  Framework independence will simplify analysis sustainability and preservation beyond the lifetime of analysis frameworks and of the experiments.  Due to its standard syntax and methodical expressions, static analysis can be applied on ADL, e.g. for analysis queries, comparisons and combinations.  Moreover, ADL can be used with differentiable programming to automate analysis design and physics model exploration.

ADL, at its current state, emerged from the combination of the best ideas from two parallel efforts: One effort is {\tt LHADA} (Les Houches Analysis Description Accord), a prototype language designed by a group of experimentalists and phenomenologists to methodically document and run content of LHC  analyses~\cite{Brooijmans:2016vro, Brooijmans:2018xbu, Brooijmans:2020yij}. The other effort is \cl~\cite{Sekmen:2018ehb, Unel:2019reo,Gokturk:2021fuu}, an initiative to build a interpreted language directly executable on events. 

\begin{figure}
    \centering
    \includegraphics[width=0.8\textwidth]{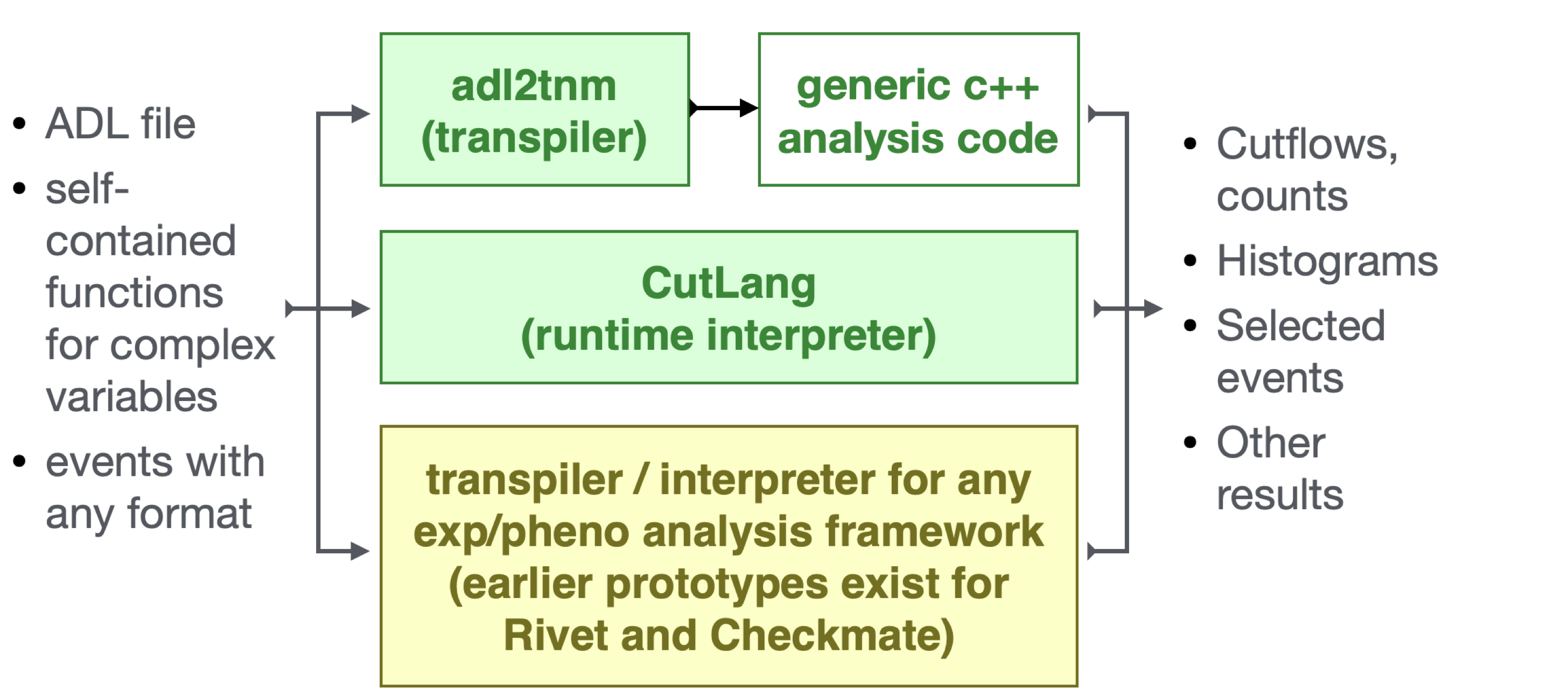}
    \caption{ADL analyses flow with different tools.  Inputs to and outputs from a typical analysis with the ADL approach.}
    \label{fig:ADLtools}
\end{figure}

ADL is in principle only a description.  Thus, for it to be practically useful, ADL must be rendered executable by tools that convert the descriptions into executable instructions.  Two approaches have been studied for this purpose: The first approach is that of direct runtime interpretation.  This is explored in CutLang, whose development started both as a language and a \emph{runtime interpreter}~\cite{Sekmen:2018ehb, Unel:2019reo, Gokturk:2021fuu}.  The second is the transpiler approach, where ADL is first converted into a GPL, which is in turn compiled into code executable on events.  A transpiler called {\tt adl2tnm} converting ADL to C++ code is currently under development~\cite{Brooijmans:2018xbu}.  Earlier prototype transpilers converting {\tt LHADA} into code snippets that could be integrated within CheckMate~\,\cite{CheckMATE,KIM2015535,2016chep.confE.120T} and Rivet~\cite{Waugh:2006ip,BUCKLEY20132803} frameworks were also studied.  All such runtime interpreter or transpiler systems are surrounded by frameworks whose purposes become reduced to handling operations such as data input, histogramming, results output for statistical analysis, etc.  Figure~\ref{fig:ADLtools} summarizes the process flow of analyses and the currently available tools.  

In this report, we will present the current status of ADL, \cl and ad2ltnm.  Section~\ref{sec:adl} will present ADL and the physics concepts it can describe.  Section~\ref{sec:cutlang} will introduce the \cl runtime interpreter and framework, along with language enhancements required for this approach, while Section~\ref{sec:adl2tnm} will introduce \atnm.    Section~\ref{sec:PhysAnalysis} will summarize the current physics implementations and uses. Section~\ref{sec:statdiff} will introduce prospects for static analysis and differentiable programming, followed by the conclusions in Section~\ref{sec:conclusions}.

\section{ADL overview: File and functions}
\label{sec:adl}

\begin{figure}
    \centering
    \includegraphics[width=0.85\textwidth]{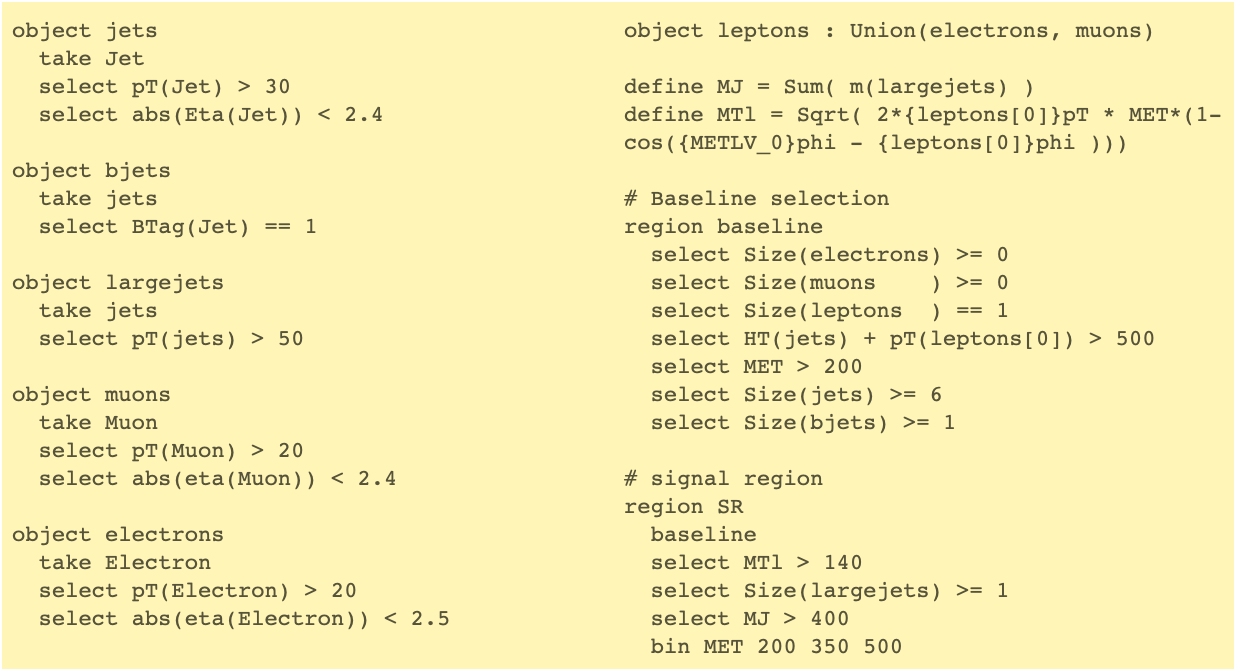} 
    \caption{An ADL example: the CMS-SUS-16-037 analysis ADL file.}
    \label{fig:example}
\end{figure}

ADL hosts the physics content of an analysis in a plain, easy-to-read text file called the {\bf ADL file}.  The ADL file consists of blocks containing one or more keyword-value/expression structures:
\begin{verbatim}
blockkeyword blockname
    keyword1 expression1
    keyword2 expression2
    keyword3 expression3 # comment 
\end{verbatim}
Blocks separate analysis components into semantically clear concepts such as object, variable and event selection definitions.  Keywords specify HEP analysis concepts and operations such as selection, weighting, binning, etc.  For example, the \texttt{select} keyword used for object or event selection is followed by a value resulting from an arbitrarily intricate boolean expression.  Tables~\ref{tab:blocks} and~\ref{tab:keywords} list the blocks and keywords currently recognized in ADL.  The syntax includes the following operators:
\begin{itemize}
\itemsep-0.4em
    \item Comparison operators: \texttt{>, <, ==, !=, =>, =<, []} (include), \texttt{][} (exclude)
    \item Optimization operators: $\sim$= (closest to) $\sim$! (furthest from)
    \item Logical operators: \texttt{and, or, not}
    \item Mathematical operators: \texttt{+, -, *, /, \^}
    \item Lorentz vector addition: \texttt{LV1 + LV2} or \texttt{LV1 LV2}.
\end{itemize}
It also includes some standard, general functions such as:
\begin{itemize}
\itemsep-0.4em
    \item Mathematical functions: \texttt{abs(), log(),} trigonometric functions
    \item Collection reducers: \texttt{size(), sum(), min(), max()}
    \item HEP-specific functions: \texttt{dR(), dphi(), deta(), m(), ….}
\end{itemize}
Sometimes analyses contain variables with complex algorithms non-trivial to express with the ADL syntax (e.g. $M_{T2}$, razor, aplanarity, etc.) or non-analytic variables (e.g. object or trigger efficiency tables, machine learning models, etc.).  ADL handles these variables systematically by having them encapsulated in self-contained, external, standalone functions that can be referenced from within an ADL file.  Throughout the ADL file, the mass, energy and momentum are all written in units of GeV and angles in radians. User comments and explanations are preceded by a hash (\#) sign. An example ADL file for a CMS analysis (CMS-SUS-16-037) is shown in Figure ~\ref{fig:example}. 

\begin{table}[h]
\caption{Blocks in ADL} 
\label{tab:keywords}
\centering{}%
\fontsize{8 pt}{1 em}
\selectfont
\begin{tabular}{|p{0.22\textwidth}|p{0.45\textwidth}|p{0.20\textwidth}|}
\hline 
Block & Purpose & Related keywords \\
\hline 
\hline 
object / obj & Object definition block. Produces an object type from an input object type by applying selections. & take, select, reject \\
\hline
region / algo & Event categorization. & select, reject, weight, bin, sort, counts, histo, save \\
\hline
info & Contains analysis information such as the experiment, center-of-mass energy, luminosity, publication details, etc. & \\
\hline
table & Generic block for tabular information, such as efficiency values versus variable ranges & tabletype, nvars, errors \\
\hline 
countformat & Expresses the processes for which external counts are included and the format of counts & process \\
\hline
\end{tabular}
\end{table}

\begin{table}[h]
\caption{Keywords in ADL}
\label{tab:blocks}
\fontsize{8 pt}{1 em}
\selectfont
\centering{}%
\begin{tabular}{|p{0.22\textwidth}|p{0.45\textwidth}|p{0.20\textwidth}|}
\hline 
Keyword & Purpose & Related block \\
\hline 
\hline 
define & Define variables, constants & -- \\
\hline
select & Select objects or events based on criteria that follow the keyword. & object, region \\
\hline
reject & Reject objects or events based on criteria that follow the keyword. & object, region \\
\hline
take / using / : & Define the mother object type & object \\
\hline 
sort & Sort an object in an ascending or descending order wrt a property. & region \\
\hline
weight & Weight events & region \\
\hline
histo & Fill histograms & region \\
\hline
process & Specify process and the format for which external counts are given & countformat \\
\hline
counts & Give external counts & region \\
\hline
tabletype & Specifies type of the table & table \\
\hline
nvars & Number of variables in a table & table \\
\hline
errors & Type of errors indicated in a table & table \\
\hline
title, experiment, id, publication, sqrtS, lumi, arXiv, hepdata, doi & Provide information about the analysis & info \\
\hline
\end{tabular}
\end{table}

ADL, in its current state, can express many standard physics tasks such as object selections based on features, basic object reconstructions, variable definitions, event selections, event weighting, etc.  However it still has some missing features. For example, ADL has no generic way to describe arbitrary combinations of objects to form new ones (e.g., the reconstruction of all possible top quark candidates from the boosted or resolved decay modes).  The prototype cannot describe low level objects (e.g. hits, cells), or non-standard objects like long-lived particles (e.g. disappearing tracks, displaced muons, etc.).  There is yet no way to add new object attributes or define object associations (e.g., between a jet and its constituent particles or a track and its associated hits).  Moreover, ADL needs to be extended with syntax to specify and apply systematic uncertainties.  Constant work is ongoing to identify and incorporate these features and evolve ADL into a domain complete language.

\section{The \cl interpreter and framework}
\label{sec:cutlang}

Runtime interpretation is a very practical approach, that allows instant modifications in an analysis such as adding new variables or selection criteria, changing the execution order or cancelling analysis steps, and evades the modify-compile-run cycle. Not having to compile ADL content into a framework  automatically provides the flexibility to run multiple analyses in parallel. \cl runtime interpreter and frameworks are developed to demonstrate the feasibility of this approach.  

\begin{figure}
    \centering
    \includegraphics[width=0.35\textwidth]{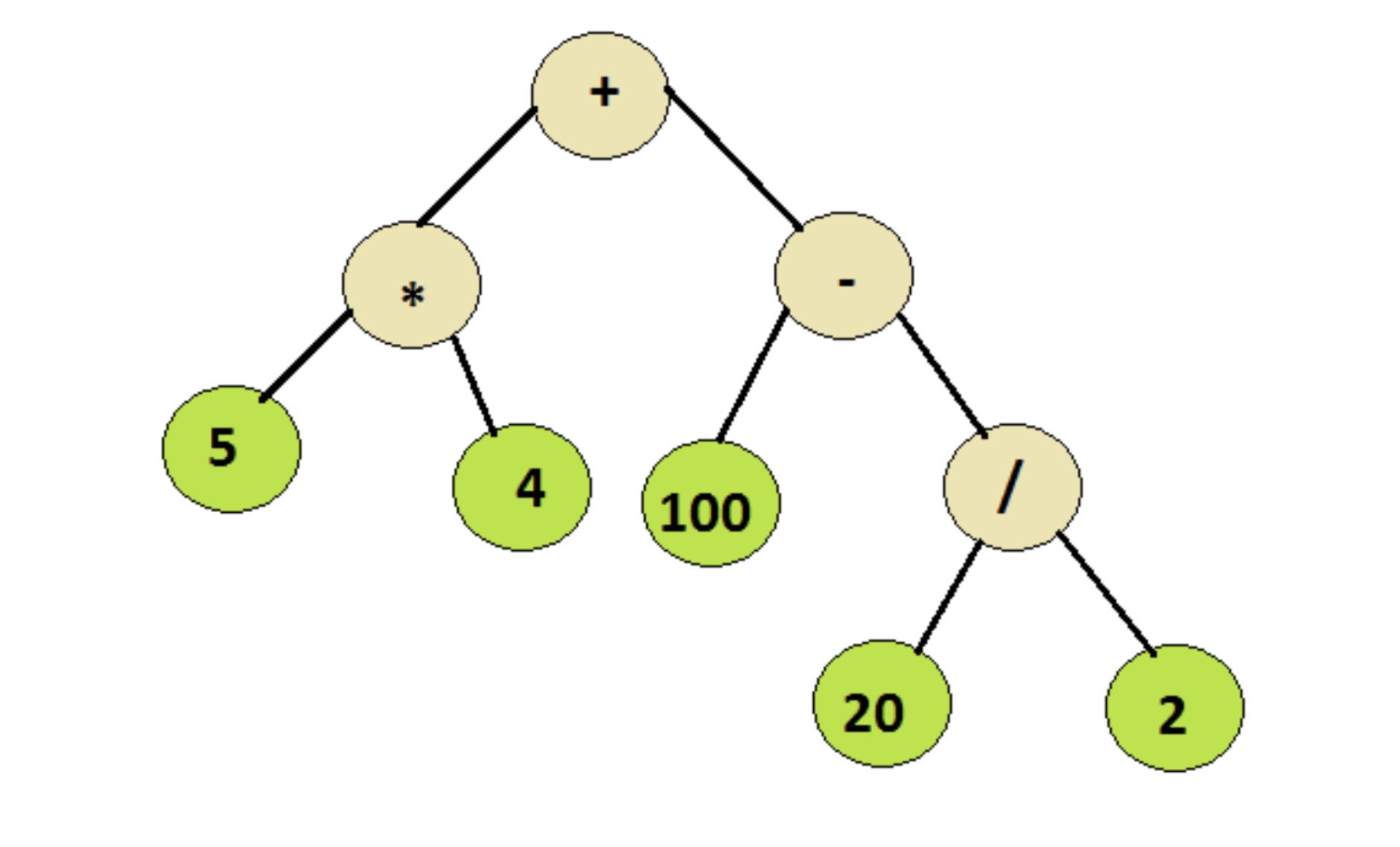} 
    \caption{An expression tree example: the program traverses the tree from right to left evaluating the encountered functions from bottom to top.}
    \label{fig:tree}
\end{figure}

\cl runtime interpreter is a  C++ program utilizing  function pointer trees to represent different operations used in event selection and other relevant functions such as filling histograms. In this approach, processing an event through a cutflow list becomes equivalent to traversing multiple expression trees, such as the one shown in Figure~\ref{fig:tree},  with arbitrary complexities. The physics objects to be used are therefore given as arguments to these functions.

Further functionalities such as handling of the Lorentz vector operations, pseudo-random number generation, input-output file and histogram manipulations are all based on classes of the ROOT data analysis framework~\cite{ROOT}. The ADL text itself is parsed in \cl, automatically to generate dictionaries and grammar using formal tools Lex and Yacc~\cite{lexyacc}. The ADL file is split into tokens by Lex, and the hierarchical structure of the algorithm is found by Yacc. Since these tools are traditionally found in all Unix-like systems, \cl can be compiled and operated in a multitude of modern Operating Systems. The interpreter is compiled only once, during the installation or if an external user function is added. Once the work environment is prepared, the remaining work consists mostly of thinking, editing, running and observing.

Multiple input data formats are implemented as plug-ins into the \cl framework.  Some of the event types that are recognized and can be directly used are ATLAS and CMS open data~\cite{opendata}, CMS NanoAOD~\cite{nanoaod}, Delphes~\cite{DELPHES} and LHCO. \cl has also its own internal format called \texttt{LVL0}.  New input file types can also be added easily: an abstraction layer defining all particle types and event properties decouples internal data from input data formats. The only requirement on the input files is to use ROOT file format.  If \cl does not provide by default the necessary methods to access some information (such as an attribute of a particle) in a particular input data type, that particular information can be easily accessed through external user functions. 

In the present design, achieving runtime interpretation inherently relies on the ADL file to comply with a certain structure and content.  For example, \cl runtime interpreter processes the commands in the ADL file on events from top to bottom.  All information, e.g. a variable name, required at a stage must be available when \cl arrives at that stage.  Therefore, in order to be processed with \cl, the description of the analysis content needs to be given in a well-defined order.  According to this order, an ADL file starts with an  \emph{initialization} section containing commands related to analysis information and initialization.  This is optionally followed by a \emph{counts} section, used for setting up the recording of already existing event counts and errors, e.g., from an experimental paper publication, if such counts are needed to be recorded for statistical analysis.  
Next section is the \emph{definitions1} used for defining aliases for objects and variables.  This is followed by the \emph{objects} section that defines new objects based on predefined physics objects and shorthand notations declared in \emph{definitions1}.  Next comes the \emph{definitions2} section used for defining more objects and variables based on all the available objects. The current implementation permits only two object definitions sections.
The final part consists of the \emph{event categorization} section that defines event selection regions, criteria in each region, event weighting and event histogramming.  \cl requires at least one selection region with at least one command, which may include either a selection criterion or a special instruction to include MC weight factors or to fill histograms. 

\cl also incorporates a complete analysis framework designed to run a full event analysis and output information and data that would be used for further study.  The main output file in the ROOT format includes a copy of the ADL file content in order to report the provenance of the analysis. The output  file contains a directory for each event  categorization region, i.e. each {\tt region} block. Under each directory, it stores histograms with the event counts and uncertainties obtained from the analysis together with all histograms filled by the user. \cl is also capable of saving the currently surviving events at any stage of the running algorithm. The events are saved into a dedicated user-defined ROOT~\cite{ROOT} file (without the .root extension) using the command \texttt{save}. It is possible to save multiple times in a single algorithm (region) at different stages of the algorithm. The events in the output file are saved in the native \texttt{LVL0} format of \cl. The ROOT file also stores the saved events in case it is declared at ADL file level. It is possible to register various signal, background or data counts of a region together with their associated errors for some studies such as phenomenological interpretation or validation.

\begin{figure}
    \centering
    \includegraphics[width=0.65\textwidth]{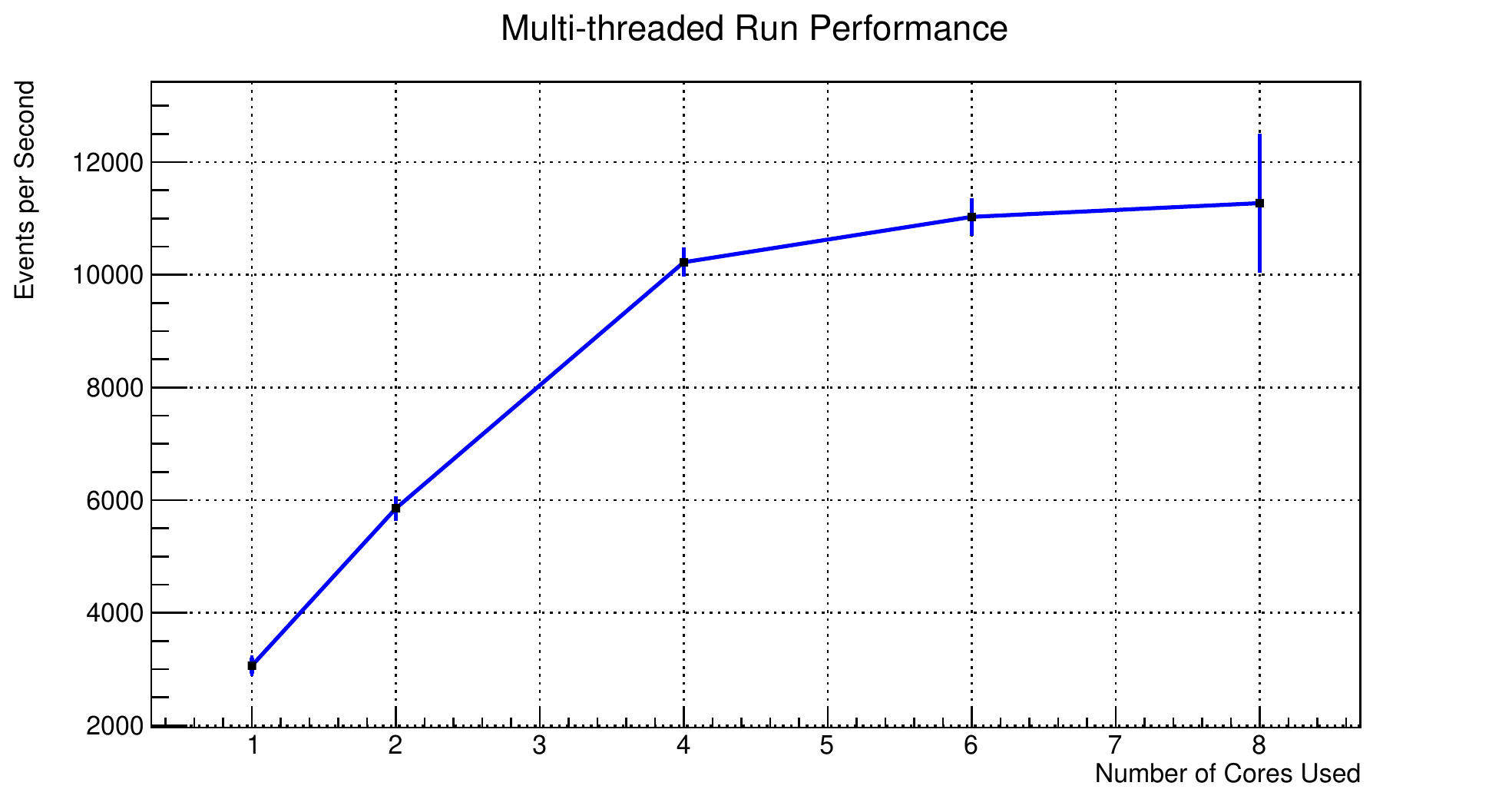} 
    \label{fig:speed}
    \caption{Multithreaded performance as a function of number of CPU cores. The linear increase changes when the system starts using the hyperthreaded cores.}
\end{figure}

\cl has also the capability of multi-threaded execution of an analysis to optimally utilize the available resources. Adding  the usual \texttt{-j n} to the command to start the analysis results in using \texttt{n} number of cores. The parameter \texttt{n} is to be an integer between \texttt{0} and total number of cores on the processor, where \texttt{0} represents a value one less than total number of cores to maximize the performance  while leaving the operating system part of the resources. The optimization choice in \cl is to parallelize over the events, which are distributed equally over the available cores.
A simple study, presented in Figure~\ref{fig:speed} shows that the optimal number of parallel processes should be equal to the number of physical cores.  Moreover, the same study showed that the multi-threading performance gets better with the increased  number of events in the analysis. This can be understood in terms of the file opening and closing overhead becoming unimportant as the total event processing time increases. As for the single thread performance, it was shown that the interpreter speed is about 20\% slower than the compiled code when used in a realistic analysis scenario~\cite{cutlang2021}.

\cl currently includes all language features explicitly listed in Section~\ref{sec:adl}.  It was tested with various physics analyses, used in one published phenomenology study~\cite{Paul:2020mul} and used as a training tool as will be described in Section~\ref{sec:PhysAnalysis}.  Yet, improvements are still needed in two areas in order for it to be usable for full scale experimental analyses.  \cl does not yet have an automated mechanism to incorporate input data formats including a complete set of objects and methods.  It also requires further automation in the incorporation of external user functions.  
\cl source code is publicly available in GitHub ~\cite{clgithub}: 

\vspace{0.1cm}
\href{https://github.com/unelg/CutLang}{https://github.com/unelg/CutLang}
\vspace{0.1cm}

Recently, the GitHub platform was  used to incorporate a continuous integration setup for  automatic validation of the code via predefined test analyses. 

\section{adl2tnm transpiler for the TNM framework}
\label{sec:adl2tnm}

\begin{figure}
    \begin{center}
    \includegraphics[width=0.65\textwidth]{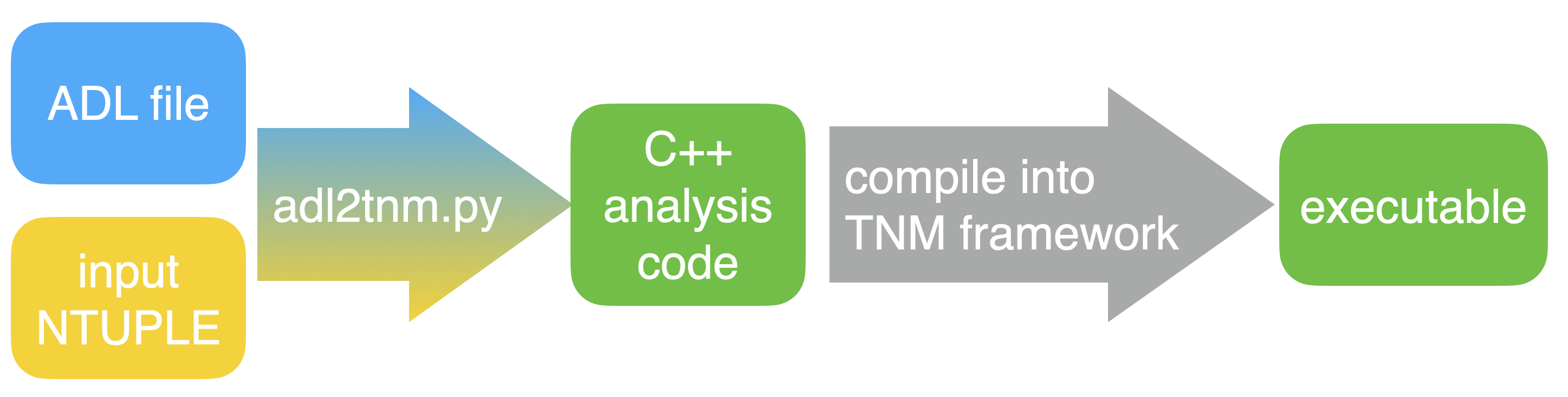} 
    \caption{The adl2tnm transpiler workflow }
    \label{fig:adl2tnm}
    \end{center}
\end{figure}

The \atnm transpiler is a Python program that translates an ADL file to a C++ program that can be executed within the {\tt TNM} ({\tt TheNtupleMaker}) framework, an automated generic ntupling and analysis framework for CMS studies.  Note, however, that the analysis component depends only on ROOT and not on any CMS data structures, therefore serving as a generic ntuple-based analysis framework. The workflow of \atnm is shown in Figure~\ref{fig:adl2tnm}.

In principle, \atnm can work with any simple ntuple event format, e.g. Delphes~\cite{DELPHES}, ATLAS and CMS analysis ntuples such as CMS NanoAOD~\cite{nanoaod}, etc.  \atnm has an adapter mechanism capable of semi-automatically reading the input event format and incorporating it into the C++ code.  \atnm operates by assuming the availability of a standard, extensible type for analysis objects, and has internally implemented such a type.  Its adapter mechanism translates the input object types to the standard extensible types.  The assumption of the standard extensible types is not an imposition on ADL itself, but rather is an aid to the writing of transpilers and interpreters for ADL.  The extensible type approach is aimed as a generic solution to handle the reality that different input types can, and do, have different attributes and sometimes identical attributes with different names. For example, the transverse momentum of a particle may be called PT, in Delphes, while the same attribute may be called Pt in other input types. Therefore, the extensible type used by \atnm uses the attribute names of the input types. The attributes are modeled as a map between a name (as a string) and a floating point value.  \atnm produces analysis output in a ROOT file with a content similar to that produced by \cl.  

\atnm does not impose an 
order within the ADL file.  The \atnm transpiler extracts all blocks from an ADL file and places them within a data structure that groups the blocks according to type. The blocks are then ordered according to their dependencies on other blocks. 

The development of \atnm started during the initial phase of LHADA.  The transpiler is not based on formal tools such as Lex \& Yacc as in the case of \cl.  Though it was tested successfully in processing several analyses in comparison to \cl, it still misses the implementation of several ADL features.  Work is in progress to re-build \atnm in a more formal way through the use of formal grammar building and parsing tools.  The current version is publicly available in GitHub~\cite{adl2tnmgithub}:

\vspace{0.1cm}
\href{https://github.com/hbprosper/adl2tnm}{https://github.com/hbprosper/adl2tnm}
\vspace{0.1cm}

\section{Physics studies}
~\label{sec:PhysAnalysis}

Up to now, various analyses, mainly from LHC new physics searches, have been implemented with ADL.  The primary goal of these implementations so far has been to determine the approximate range of physics content and design ADL syntax to address this content.  Implementing analyses with a variety of physics content led to incorporating a wider range of object and selection operations and helped to make the ADL syntax more generic and inclusive. Consequently, the scope and functionality of \cl interpreter and \atnm transpiler and frameworks were also significantly enhanced.  These ADL analyses are being collected in the following GitHub repository~\cite{adllhcanl}:

\vspace{0.2cm}
\href{https://github.com/ADL4HEP/ADLLHCanalyses}{https://github.com/ADL4HEP/ADLLHCanalyses}
\vspace{0.2cm}

These ADL implementations have been tested with \cl and partially with \atnm.  Some of them were also validated in comparison to other analysis frameworks in dedicated exercises performed during the Les Houches PhysTeV workshops, ~\cite{Brooijmans:2020yij} (Contribution 19).  The phenomenology tool  \texttt{SModelS}\,\cite{Kraml:2013mwa,Ambrogi:2017neo,Ambrogi:2018ujg} which decomposes a given new physics model into a set of simplified final states, and uses the experimental limits from various analyses on these simplified final states to obtain the sensitivity to the model is adapting ADL and \cl to compute the analysis selection efficiencies of the simplified model final states.  

More recently, ADL and \cl were used in a study estimating the sensitivity of the High Luminosity LHC and the Future Circular Collider to models with  down-type isosinglet quarks~\cite{Paul:2020mul}. Furthermore, an analysis example to run on CMS Open Data~\cite{opendata} was implemented.  In addition, ADL and \cl were used as main tools in an analysis school which took place in Istanbul in February 2020 for undergraduate students, where several analyses were implemented by the participating students~\cite{Adiguzel:2020brl}.  ADL and \cl were also employed in hands-on exercises for data analysis at the 26th Vietnam School of Physics (VSOP) in December 2020~\cite{vsop}, where the exercises were adapted to be performed via Jupyter notebooks~\cite{vsophandson}.  The experience in both schools established ADL and \cl as highly intuitive tools for introducing HEP data analysis to beginner level students.

\section{Prospects for static analysis and differentiable programming}
\label{sec:statdiff}

The formal domain specific, declarative syntax and the well-defined structure of ADL makes it an ideal construct for implementing static analysis and differentiable programming.  Moreover, having the analysis described in an independent text file decoupled from framework code greatly aids such tasks.

The act of parsing source code and deducing facts about it without actually running the code is called a \emph{static analysis}.  Static analysis of a database of physics analyses implemented with ADL can be used  to assist and automate query among or comparison between multiple analyses in the space of event properties. This helps to find out which event final states are covered or not, and which analyses have disjoint or overlapping selection regions. The practical features of ADL makes such comparison tasks possible to some extent ``by eye'', even without formal static analysis.  In any case, this information can in turn be used to combine multiple analyses or design original ones.  Recently, we started to develop prototype tools for analysis queries and comparisons.  The tools are designed to have various options to perform these tasks, i.e. via static analysis, via using physics events or via using randomly sampled events. A more detailed description of these methods can be found in Contribution 17 of~\cite{Brooijmans:2020yij}, and a preliminary version of the tools can found in the repository~\cite{adl2tnmgithub}.  

The task of a HEP data analysis can be viewed as a mathematical function, which takes as arguments signal, background and observed events, various cross sections, and interfaces with a statistical tool providing a desired output, such as a measure of statistical significance for a sought signal or a measurement (e.g., of a cross section) and its associated uncertainty. The mapping from events to desired outputs is an optimization problem. For example, in the case of expected statistical significance, the goal is to maximize it. For a measurement, the goal may be to reduce the expected relative measurement uncertainty. Therefore, a HEP data analysis fits into the broad class of optimization problems whose solution is, in principle, amenable to optimization using gradient descent or ascent. Such problems may be effectively handled via differentiable programming where the analysis elements such as selection thresholds are treated as differentiable parameters. The ADL approach is particularly suited to this approach as it uniquely and systematically organizes the description of these parameters.  A dedicated effort has started in the HEP community towards building automatic differentiation tools to make analyses completely differentiable and in particular developing differentiable replacement analogues for non-differentiable operations such as binning and sorting that are common to HEP analyses~\cite{gradhep}.  ADL will be combined with these emerging tools to obtain differentiable analyses.


\section{Conclusions and outlook}
\label{sec:conclusions}

In this paper, we presented the concept and recent developments in a domain specific, declarative and framework-independent Analysis Description Language for HEP analyses.  We gave an overview of the current ADL syntax, which is already reasonably sufficient for describing the physics content of a large number of analyses.  We then presented the two main tools developed to render ADL executable, namely the runtime interpreter \cl and the transpiler \atnm.  Both tools can already be used for processing various analyses on events and produce meaningful output that can be used in further statistical studies.  Currently, \cl supports a wider range of ADL features while \atnm has a more automated way of handling input data formats and external functions.  We also discussed the prospects of ADL for statistical analysis and differentiable programming and presented the existing and ongoing physics applications.  All these studies  demonstrate the feasibility, effectiveness and potential of ADL, and establish motivation to  pursue this initiative and its diverse applications.  Up-to-date information about ADL, \cl , \atnm and various applications is systematically documented at the project's web portal \href{cern.ch/adl}{cern.ch/adl}~\cite{adlweb}. Studies will continue towards developing ADL into a domain complete language, improving the functionality and robustness of \cl and \atnm, to build new tools making use of ADL's potential and practicality, and to explore a large variety of physics applications.

\section*{Acknowledgements}

SS is supported by the National Research Foundation of Korea (NRF), funded by the Ministry of Science \& ICT under contracts NRF-2008-00460 and 2021R1I1A3048138.
This work was supported in part by the US Department of Energy Award No.DE-SC0010102

\bibliography{refs}

\end{document}